\documentclass[12pt]{article}
\usepackage{epsfig}
\usepackage{epsfig}
\begin{document}
\def\l{{(\lambda)}}
\def\tr{{\rm tr}\, }
\def\Tr{{\rm Tr}\, }
\def\hTr{\hat{\rm T}{\rm r}\, }
\def\be{\begin{eqnarray}}
\def\ee{\end{eqnarray}}
\def\ctt{\chi_{\tau\tau}}
\def\cta{\chi_{\tau a}}
\def\ctb{\chi_{\tau b}}
\def\cab{\chi_{ab}}
\def\cba{\chi_{ba}}
\def\ptt{\phi_{\tau\tau}}
\def\pta{\phi_{\tau a}}
\def\ptb{\phi_{\tau b}}
\def\>{\rangle}
\def\<{\langle}
\def\d{\hbox{d}}
\def\pab{\phi_{ab}}
\def\lb{\label}\def\appendix{{\newpage\section*{Appendix}}\let\appendix\section%
        {\setcounter{section}{0}
        \gdef\thesection{\Alph{section}}}\section}
\renewcommand{\figurename}{Fig.}
\renewcommand\theequation{\thesection.\arabic{equation}}
\hfill{\tt  IHES/P/07/19}\\\mbox{}
 \vskip0.3truecm
\begin{center}
\vskip 2truecm {\Large\bf Wormholes as Black Hole Foils}
\vskip 1.5truecm {\large\bf Thibault Damour\,${}^*$ and Sergey N.
Solodukhin\,${}^{*\dagger}$}
 \vskip 0.6truecm \noindent ${}^*${\it
 Institut des Hautes Etudes Scientifiques, \\ 35, route de
 Chartres,\\
 91440
  Bures-sur-Yvette, France
  }\\
\vskip 0.6truecm
 \noindent ${}^{\dagger}${\it School of Engineering and Science,\\ International
 University Bremen,\\
   Bremen  28759, Germany}
\end{center}
\vskip 1cm
\begin{abstract}
\noindent We study to what extent wormholes can mimic the
observational features of black holes. It is surprisingly found
that many features that could be thought of as ``characteristic''
of a black hole (endowed with an event horizon) can be closely
mimicked by a globally static wormhole, having no event horizon.
This is the case for: the apparently irreversible accretion of
matter down a hole, no-hair properties, quasi-normal-mode ringing,
and even the dissipative properties of black hole horizons, such
as a finite surface resistivity equal to 377 Ohms. The only way to
distinguish  the two geometries on an
observationally reasonable time scale would be through the
detection of Hawking's radiation, which is, however, too weak to
be of practical relevance for astrophysical black holes. We point
out the existence of an interesting spectrum of  quantum
microstates trapped in the throat of a wormhole which could be
relevant for storing the information ``lost'' during a gravitational
collapse.
\end{abstract}
\vskip 1cm
\newpage

\section{Introduction}
\setcounter{equation}0
 One of the most striking predictions of
Einstein's theory of gravity is the existence of black holes.
Though these objects made their first appearance in the famous
exact spherically symmetric solution found by Karl Schwarzschild
\cite{Schwarzschild:1916uq} a couple of months after Einstein
finalized his theory, it took many years, and the work of many
physicists, to cristallize the concept of black hole (see,
e.g.,\cite{Misner:1974qy}). For a long time, a part of the physics
community was rather sceptical about the actual existence of black
holes, but the situation has changed in recent years, notably
because of several different types of astronomical observations:
in X-ray binary systems, in galactic nuclei (including our home,
the Milky Way), etc. For a review of the astronomical evidence for
black holes see \cite{Narayan:2005ie}.

Today black holes are  part of the basic ``toolkit'' of physicists
and astrophysicists, and their existence in the real universe is
taken for granted. It is, however, interesting to examine
critically to what extent the current, or future, astrophysical
data can {\it observationally  prove} the existence of black
holes. Indeed, black holes are sophisticated theoretical
constructs with many different properties and each observational
evidence usually concerns only one specific property. For
instance, in many observations  the ``black hole candidates''  are
mainly picked either because their inferred mass exceeds some
theoretical limit, or on the basis of their  strong external
gravitational field. Several authors have claimed that some
observations have probed, or will eventually probe, more
characteristic features of black holes, and notably the
(essentially defining) existence of an event horizon. For
instance, Narayan and collaborators have argued that, in several
examples, a black hole candidate ``does not have a surface, i.e.
it must have an event horizon''
\cite{Narayan:2005ie,Broderick:2005xa}. In a different vein, it is
also commonly argued that forthcoming gravitational wave data from
LIGO/Virgo/GEO will establish the existence and ``unique''
properties of black holes either through the observation of the
characteristic ``quasi-normal mode'' (QNM) ringing frequencies of
a newly formed hole \cite{Kokkotas:1999bd}, or from observational
checks of the ``unique'' structure of the black hole geometry
guaranteed by ``no-hair'' theorems
\cite{Ryan:1995wh,Hughes:2005wj}.

A well-known, and useful strategy for gauging the extent to which
observations can really characterize the presence of  general
relativistic black holes is to consider ``black hole foils'', i.e.
theoretical objects that mimic some aspects of black holes, while
lacking some of their defining features. Several examples of this
strategy have been considered in the past. For instance, would-be
black holes within Rosen's bimetric theory of gravity
\cite{Stoeger:1978hp}, or, more recently, some ``gravstar'' models
\cite{Broderick:2007ek}.

In this note, we consider a very simple type of black hole foil: a
wormhole \cite{Visser:1995cc}. Though a wormhole does not have an
event horizon, and differs, in principle,
 in several other important ways from a black hole, we shall show
 here that, if a certain parameter entering its definition
  is small enough, a wormhole is essentially astrophysically
  indistinguishable from a black hole. Our final conclusion
is that the possibly unique way of conclusively proving the
presence of a black hole (endowed with an horizon) would be to
observe its Hawking radiation \cite{Hawking:1974sw}. And even this
conclusion needs some qualification, because we shall see that
some features of wormholes naturally tend to mimic the quantum
spectrum of black holes, so that it is possible that some (to be
defined) wormhole formation mechanisms could lead to an
Hawking-like radiation.

\section{ Wormhole metric}
\setcounter{equation}0
 We shall consider here a very simple type
of wormhole spacetime, as described by the metric
 \be ds^2=-(g(r)+\lambda^2)dt^2+{dr^2\over
g(r)}+r^2(d\theta^2+\sin^2\theta d\phi^2)\, , \lb{1} \ee
 where
$g(r) \equiv 1-{2GM\over r}$. This metric differs from the
standard Schwarzschild metric \cite{Schwarzschild:1916uq} only
through the presence of the dimensionless parameter $\lambda$.
When $\lambda = 0$ we recover
 a black hole of
mass $M$ with an event horizon located at the radius $r=2GM$. By
contrast, when $\lambda\neq 0$
 the structure of the spacetime is
dramatically different: there is no event horizon, instead there
is a throat at $r=2GM$ that joins two isometric, asymptotically
flat regions. This spacetime is an example of  a Lorentzian
wormhole \cite{Visser:1995cc}. In three dimensions a similar
modification of the black hole metric was studied in
\cite{Solodukhin:2005qy} in an attempt to restore  Poincar\'e
recurrences in black holes. The parameter $\lambda$ in the latter
construction was chosen to be exponentially small \be \lambda\sim
e^{-4\pi GM^2} \lb{2} \ee in order to reproduce the expected
dependence of the Poincar\'e recurrence time on the entropy of a
black hole. Though we shall leave free the value of $\lambda$ in
this paper, and discuss what range of values for $\lambda$ is
compatible with present and foreseeable observations, we will see
below that exponentially small values of the type of (\ref{2})
seem indeed adequate for mimicking not only the classical, but
also the quantum properties of  a Schwarzschild black hole.

The event horizon of the original black hole metric is replaced,
in the wormhole metric (\ref{1}), by a high-tension distribution (a
kind of {\it brane}) localized in a thin shell around the center
of the {\it throat} at $r=2GM$. More precisely, with our
simplifying choice of wormhole metric (\ref{1}), the stress-energy
tensor distribution sustaining the throat has vanishing energy
density, but comprises radial and tangential tensions proportional
to $1/\lambda^2$.

In order to define globally the wormhole spacetime (\ref{2}) we
need to specify how the spacetime is continued through the
(geometrically regular) throat $r = 2GM$. The Schwarzschild-type
radial coordinate $r$ is not well defined at $r = 2GM$. We should
replace it, for instance, by the proper radial distance, say
 $y=\int_{2GM}^r dr /\sqrt{g}$. In terms of $y$, one has, to leading order, the following
expressions in the throat: $g(y)={y^2\over 16G^2M^2}$ and
$r(y)=2GM+{y^2\over 8GM}$. Using the coordinate $y$ we can now globally
define the wormhole spacetime in several different (physically
inequivalent) ways. A first possibility (which is the usual one
when considering ``wormholes'') is to decide that the variable $y$
varies over the full real line: $-\infty<y<+\infty$. A second
possibility is to impose some $Z_2$ symmetry between $y$ and
$-y$, so that
 $y$ effectively varies only on a half-line $ 0 \leq y<+\infty$ (with some
 $Z_2$-symmetry boundary conditions at $y=0$). We might prefer the
 first possibility if we have in mind a multi-brane-world in which the
 collapse of a star establishes a bridge between two previously separate
 brane-worlds. If, instead, we have in mind a unique world, we might prefer
 imposing the second possibility, i.e. the idea that the collapse of a
 star creates an ``end-of-the-world'' $Z_2$-symmetric brane at $r = 2GM$,
 which is certainly a logically allowed possibility.
Note then that, in both cases, the wormhole spacetime (\ref{2}) is
globally static, the time Killing vector being everywhere timelike
(while it became spacelike beyond the horizon in the black hole
case).

An immediate consequence of the metric (\ref{1}) is that  time in the
throat is extremely slow from the point of view of a distant
observer. Indeed, they are related by $\lambda$,
$$
t_{\rm thr}=\lambda t_{\rm dist}~~.\lb{2a}
$$
The throat thus mimics what happens at the event horizon of a
black hole
 where time is ``frozen" [we recall that the old name (especially in Russia)
 for a black hole was a ``frozen star''].
The only difference from an actual horizon is that time does not
completely stop in the throat: if an observer makes observations
during a time of order $ G M/\lambda$ he or she will resolve the
processes happening in the throat and thus be able to distinguish
a wormhole from a black hole. Reciprocally, this preliminary
remark suggests that if an observer only looks at a wormhole
during a finite time  he or she might not be able to distinguish
it from a black hole. We shall see below, in several examples,
that this is indeed the case, even for phenomena that are usually
considered as characteristically linked to the presence of an
horizon (such as no-hair properties, or dissipative properties).
However, we shall see that the observing time span needed to
distinguish a wormhole from a black hole is not $ G M/\lambda$, as
suggested by the above naive argument,  but rather $ G M/ \ln
(1/\lambda)$.

\section{ Geodesics}
\setcounter{equation}0
 As first, and simplest example of the comparative phenomenology
 of wormholes versus black holes, let us consider the motion of
 particles around a wormhole, and their fall within the throat.

 The (equatorial) geodesics in the metric (\ref{1}) are
 described by the equations
\be &&\dot{t}={E\over g(r)+\lambda^2} \, , \, \ \
\dot{\phi}={L\over
r^2}~~, \nonumber\\
&&\dot{r}^2+g(r)({L^2\over r^2}+\epsilon)={g(r)\over
g(r)+\lambda^2} E^2~~, \lb{3}\ee where $\epsilon=0$ for a
null-like geodesic, $\epsilon=1$ for a time-like geodesic, and
where the overdot denotes the derivative with respect to the
proper time (or an affine parameter, in the null case). $E$ is the
energy and $L$ the angular momentum of the test particle (for
simplicity we consider a
 test particle of unit mass). In terms of the new
coordinate $\rho=\int \sqrt{g+\lambda^2\over g}dr$ the last
equation in (\ref{3}) takes the standard form
\be &&\dot{\rho}^2+V(r(\rho))=E^2~~,\nonumber \\
&&V(r(\rho))=(g(r)+\lambda^2)({L^2\over r^2}+\epsilon ) ~~.\lb{4}
\ee The consideration of the ``effective potential'' $V(r)$ (or
rather $V(r(y))$ to understand what happens in the throat) then
allows one to understand qualitatively the dynamics of particles
in the wormhole.

As soon as one is a little bit away from the throat, the dynamics
is that for the Schwarzschild metric plus (observationally
negligible) small corrections proportional to $\lambda^2$. This
shows that any observational feature which is not taking place
very near $r=2GM$ will be (for small enough $\lambda$) the same in
the wormhole foil than in a real black hole. This is for instance
the case for the emissivity properties of accretion disks, even
those that crucially depend on relativistic dynamics features
(like the presence of an innermost stable circular orbit (ISCO)).

On the other hand, there is an  important difference from the
black hole case if we consider, say, circular
 orbits with radius  equal to (or very near) the throat radius $r=2GM$.
 Both for null and timelike
geodesics, and  {\it for any value of the angular momentum
$L$},\footnote{including $L=0$, in which case one has an
equilibrium position at fixed $r$ and $\phi$.} there exists a
circular orbit exactly located at $r=2GM$. The energy and angular
momentum in this case are related by $E^2=\lambda^2({L^2\over
4G^2M^2}+\epsilon)$. The derivative of the radial potential
vanishes $\partial_\rho V=0$ for $r=2GM$ while the second
derivative $\partial_\rho^2 V(\rho))={1\over
  2\lambda^2}V'_rg'_r|_{r=2GM}$ is positive. Thus, this ``throat-orbiting''
circular orbit is stable. In addition, the positive curvature of
the effective potential at $r=2GM$ implies that there exist bound
``elliptic'' orbits staying near $r=2GM$.

An argument often evoked for distinguishing a black hole from
other potential wells is the absence of a ``surface'' in the black
hole case, and the possibility for the horizon of absorbing any
amount of infalling matter \cite{Narayan:2005ie,Broderick:2005xa}.
The situation is a priori quite different in the wormhole case
because a look at the qualitative shape of the effective potential
$V(r(y))$ shows that, for instance,  particles falling from (just
below) the ISCO must ultimately bounce back up again from the
wormhole throat.\footnote{This conclusion holds in the case where
$y$ ranges over the full real line (in which case the effective
potential is made of two $y$-symmetric humps), as well as in the
$Z_2$-symmetric case where the effective potential has only one
hump, but where the particle bounces off when it reaches $y=0$.}

To study in more detail this ``bounce'' from a wormhole, let us
focus for simplicity on the case of  radial timelike geodesics,
with $L=0$ and therefore $V(r) = \epsilon (g(r)+\lambda^2) $ (with
$\epsilon=1$). There are 3 cases. If $E^2>1+\lambda^2$ then a
particle coming from infinity falls into the wormhole. These geodesics
are similar to the radial geodesics in the Schwarzschild metric.
If $\lambda^2<E^2<1+\lambda^2$ then the geodesic has a turning
point $y=y_m$ in which $\dot{y}=0$. The coordinate of the turning
point must solve the equation $g(y_m)+\lambda^2=E^2$. There are
exactly two  solutions to this turning-point equation, a positive
one $y_m >0$ and its opposite $- y_m <0$. If the wormhole connects
two separate spaces, these turning points are physically distinct,
and the radial geodesic bounces back, in an oscillatory manner,
between them. In the $Z_2$-symmetric case, the radial geodesic
bounces between the positive turning point and $y=0$ (with half
the period taken in the former case).
 There is no analog of
these geodesics in the case of the Schwarzschild metric. Finally,
if $E^2<\lambda^2$ there is no solution to the geodesic equation.
Note also that for null radial geodesics( $L =0$, $\epsilon=0$, so
that the effective potential {\it vanishes}) there are no
oscillating solutions in the two-separate-spaces case, the light
irreversibly falls into the wormhole as it does in the black hole.
But, in the $Z_2$-symmetric case, light bounces back at the throat
and exits from the mouth of the wormhole.

All this suggests that present observations of accreting (or
formerly accreting) gravitational potential wells rule out their
modelling as wormholes. However, as anticipated above from the
basic  scaling (\ref{2a}) it is crucial to study on what time
scale the matter which falls within a wormhole does come out again
in our universe. Let us compute  the coordinate time taken by a
particle ($\epsilon=1$) to go (by geodesic motion) from a point
outside the wormhole, $y=l>0$ to a point inside the throat (say
$y=0$). The same calculation, but now taken for $\epsilon=0$, will
give the coordinate time taken by a light signal to join a point
inside the throat (say $y=0$) to a point outside the wormhole
$y=l>0$. These coordinate times are given by

\be &&\Delta t=\int_0^l{Edy\over
\sqrt{(g+\lambda^2)(E^2-\lambda^2-g)}} \, ,\
\epsilon=1 \nonumber \\
&&\Delta t=\int_0^l{dy\over \sqrt{g+\lambda^2}} \, , \ \ \ \ \ \ \
\ \ \ \ \ \ \ \ \ \ \ \ \ \ ~ \epsilon=0 \lb{5} \ee Irrespectively
of the type of the geodesic this time is dominated by the throat
region, i.e. the part of the integrals where $ g(y) \sim
\lambda^2$. As both integrals are logarithmically divergent in the
black hole limit, $\lambda \to 0$, $ \Delta t \sim \int
dy/\sqrt{g} \sim \int dr/g(r)$, it is easily seen that
 the leading term when $\lambda \neq 0$ is
 \be \Delta t= 2 GM \ln{1\over
\lambda^2}~~. \lb{6}  \ee This result shows that if $\lambda$ is
small enough, it is impossible for observations extending on some
limited time scale $T$ to distinguish the provisory fall of matter
in a wormhole from the irreversible absorption of matter down the
horizon of a black hole. For instance, if we consider the
candidate ``massive black hole'' ( with $M \sim 3 \times 10^6
M_{\odot}$) at the center of our Galaxy, and assume it started
accreting matter $6$ billion years ago, it could be a wormhole if
$\lambda \ll \exp (- 10^{15})$.

We clearly need exponentially small values of $\lambda$ to mimic
observational facts. Note that this would be precisely the case if
we were using the value (\ref{2}) above, as suggested for quite
different reasons in \cite{Solodukhin:2005qy}. Actually, if we
substitute here the value (\ref{2}) for the parameter $\lambda$ in
the ``wormhole bounce time scale'' (\ref{6}) we get \be \Delta
t=16\pi G^2M^3 \lb{7}\ee which is of the order of the Hawking
evaporation time scale for a Schwarzschild black hole. We shall
come back below to this suggestive link between quantum black
holes and (classical and/or quantum) wormholes.

Before discussing other phenomenological aspects of wormholes, let
us mention a potential difficulty with the wormhole model proposed
here. In this paper we shall content ourselves with a first-order
treatment in which the matter and fields ``falling into'' a
wormhole are considered as test-matter propagating in a given
wormhole background. However, the stress-energy tensor carried by
all the matter that have accreted in the past onto a wormhole (and
that is, for all practical purposes,
 essentially ``frozen'' around the throat $r=2GM$) will distort the
 background wormhole metric. However, it is well known \cite{Visser:1995cc}
 that a wormhole requires that the ``matter'' making it up must
 violate (some form of) the positive-energy condition (we saw that above in the
 fact that the energy density corresponding to the metric (\ref{1}) vanishes,
 while the tension does not vanish). As the accreted
 matter does satisfy the usual positive energy conditions, it is not clear
 how much accreted matter can be allowed in before risking to destroy
 the wormhole throat. Actually, we should provide a more complete definition of
our wormholes as {\it dynamical objects}. For instance, one
should, in principle, discuss the dynamical structure of the
``brane'' located at $y=0$, and its possible interaction with the
matter falling onto it. Even without such a complete dynamical
definition, we think that it is interesting to explore, as we do
here, how wormholes can be conceptually clarifying foils for black
hole dynamics.

\section{ No-hair properties}
\setcounter{equation}0
 As an example of the way
wormholes can mimic the no-hair properties of black holes, let us
consider static axisymmetric (test) electric fields in a general
wormhole background $ ds^2 = - A(r) dt^2 + B(r) dr^2 + r^2
(d\theta^2+\sin^2\theta d\phi^2)\,.$
 The static Maxwell equations ,
$ \partial_\nu (\sqrt{-g} F^{\mu \nu}) =0$ reduce (when taking
$\mu = t$) to a second-order differential equation (in $r$ and
$\theta$) for the electric potential $A_t$. One easily separates
the $r$ and $\theta$ dependence by factoring: $A_t(r, \theta) =
a(r) P_l(\cos \theta)$, where $P_l(\cos \theta)$ is a usual
Legendre polynomial. This leads to the following separated
equation for the radial factor $a(r)$ \be \sqrt{A \over B}
\partial_r ( {r^2 \over \sqrt{AB}} \partial_r a(r)) = l(l+1)
a(r)~~. \ee Let us consider, for example, the wormhole metric $A=
g + \lambda^2, B=1/g$, taken in the $Z_2$-symmetric case. One
generically sees that a solution which is regular and
$Z_2$-symmetric in the throat ( $ da/dy = 0$ at $y=0$) will grow
like $r^l$ at infinity. Therefore we indeed have a no-hair
property paralleling the one for black holes: the only solution
which is regular at the throat, and decaying at infinity, is the
trivial one $a(r) = 0$ for any $l$ \footnote{In the
non-$Z_2$-symmetric case, the monopolar case, $l=0$, is a special
case in that there exists a source-free everywhere regular
solution parametrized by a charge $Q$, namely  $\partial_y a(r)= {
{Q \sqrt{A}}  \over r^2} $.}.

{}From this no-hair property one deduces that if one brings a
point charge  near the throat $r=2GM$ (but parametrically far away
from the $r-2GM \sim 2GM \lambda^2$ ``near-throat limit'') this will
generate an electric field which is essentially indistinguishable
(modulo corrections $\propto \lambda^2$) from the one generated
near a black hole, i.e. an electric field which, when seen from
outside,
 erases the information about the
location of the point charge and  looks like a spherically
symmetric electric field centered on the hole (see
\cite{Hanni:1973fn}).

\section{ Quasi-normal mode ringing}
\setcounter{equation}0
 It is often said that the
observation by the LIGO/Virgo/GEO network of gravitational wave
detectors of the quasi-normal mode (QNM) ringing of a newly formed
(rotating) black hole will provide an excellent confirmation of
the actual existence of black holes in Nature
\cite{Kokkotas:1999bd}. Indeed, the definition of QNM modes
depends in a crucial way on the presence of an event horizon. Let
us recall that the QNM modes are defined as complex-frequency
eigenmodes which satisfy the boundary conditions of being {\it
outgoing at radial infinity}, and {\it ingoing towards the black
hole horizon}. To discuss what happens of QNM modes in a wormhole
background let us consider, for simplicity, the case of {\it
scalar field} modes. [Our physical discussion will make it clear
that our conclusions apply to the more relevant {\it (tensor)
gravitational} excitations.]

For a mode of a scalar field of frequency $\omega$, $\Phi={1\over
r}\psi(r) e^{-i\omega t}Y_{lm}(\theta,\phi)$, we get a radial
equation in the Regge-Wheeler-Zerilli form, i.e. an
 effective Schroedinger equation \be
&&-\psi_{zz}+U(z)\psi=\omega^2\psi \, , \nonumber \\
&& U(z)={r_{zz}\over r}+{l(l+1)\over r^2}(g(r)+\lambda^2)~~.
\lb{8} \ee Here, a $z$ subscript denote a $z$ derivative,
 $l=0,\ 1,\ 2, \dots$ and we have used as variable the
``tortoise'' radial coordinate $z= \int_{2GM} dr \sqrt{B \over A}=
\int_{2GM} {dr\over \sqrt{g(g+\lambda^2)}}$, which is usually
denoted $r_{*}$, and which is such that the radial part of the
metric ($ds^2 = - A dt^2 + B dr^2$) is conformal to $-dt^2 +
dz^2$.
 Inserting $r_{zz}=g'_r(g+{\lambda^2\over 2})$, we get the
 following explicit form for the effective
radial potential
 \be &&U(r)={2GM\over
r^3}(1+{\lambda^2\over 2}-{2GM\over r}) +{l(l+1)\over
r^2}(1-{2GM\over r}+\lambda^2)~~. \lb{9} \ee For a wormhole
connecting two separate spaces, the tortoise radial coordinate $z$
varies over the full real line, and this potential is made of two
separate positive humps located at the positive and negative
values of $z$ corresponding to $r \sim 3GM$. In the
$Z_2$-symmetric case we have only one hump, together with a
suitable boundary condition at the throat located at $z=0$. Let us
think in terms of the easily visualizable two-humped potential.
[The $Z_2$-symmetric case consists anyway in retaining the
$z$-even solutions of the other case.]

This two-humped potential clearly has a very different spectrum
than the usual black hole effective potential which has only one
positive hump located around $r \sim 3GM$. If we look for modes
satisfying the usual QNM condition of being purely outgoing (both
towards $ z\to + \infty$ and towards $ z\to - \infty$) we will
have a spectrum which is qualitatively very different from the
usual black hole QNM spectrum. Indeed, it will now contain modes
with a real part of the frequency  lower than the maximum of the
effective potential, and a very small imaginary part. These modes
are quasi-bound states (``resonances'') trapped within the
two-humped potential, with a small escape probability and a long
lifetime. There is no analog of these modes in the black hole
case. As for the former black hole QNM modes, they do not seem to
play any prominent role anymore. Indeed, near, but  on the left of
the rightmost hump, there will exist, for a general ``wormhole QNM
mode'', a combination of left-moving and right-moving modes which
has nothing to do with the black hole QNM modes which are purely
left-moving (i.e. away from the summit of the potential).

Have we got here a clear observable distinction between a wormhole
and a black hole? In fact not. Indeed, the observable way in which
one hopes to detect QNM ringing in the black hole case consists in
considering the signal emitted by a source falling into the hole,
i.e. a source starting at some large and positive value of $z$,
and  moving leftwards towards smaller values of $z$. The
observable signal emitted by this source is obtained, in the time
domain, by the convolution of the retarded Green function $G_{\rm
ret}(t-t',z,z')$ corresponding to the Klein-Gordon-like
(time-domain) equation (\ref{8}) with the source, say $\delta(z' -
z_{\rm geodesic}(t'))$. It is true that the wormhole retarded
Green function $G_{\rm ret}(t-t',z,z')$ is globally quite
different in the wormhole spacetime, compared to the black hole
case, because there will be a combined diffusion effect due to the
two potential humps. However, if the observer looks at the emitted
signal only over time scales much smaller than the time it takes
for a causal signal to go from a source event $(t',z_{\rm
geodesic}(t'))$ [located, say, near the rightmost hump] to the
leftmost hump, and then to scatter back to the right until the
observation event $(t,z)$, the observed signal will be the same as
that computed by using only the diffusion effect of the rightmost
hump, i.e. the retarded Green function of a black hole
spacetime. And this computation, when done in the Fourier domain,
will exhibit phenomena linked to the usual black hole QNM modes.
This indirect, but physically clear reasoning, shows that if
$\lambda$ is such that the time scale (\ref{6}) is longer than the
observational time scale the signals emitted by a source falling
into a wormhole will contain the usual QNM ringing ``signature of
a black hole'', in spite of the absence of a true horizon in the
wormhole case.

\section{ Dissipative properties}
\setcounter{equation}0
 Let us now discuss whether
wormholes can mimic the dissipative properties of black holes,
and notably the fact that they can be described as membranes
having a finite electric (surface) resistivity equal to 377 Ohms
\cite{Damour:1978cg} (as well as a finite (surface) viscosity
\cite{DamourMG2}). Indeed, the proof of these properties crucially
relies on the presence of an horizon.

We might have here a good way of observationally discriminating
wormholes from black holes. For instance, we can consider the
physical situation discussed in \cite{Damour:1978cg}. An electric
current $I$ is passed through a black hole, penetrating through
the North pole and exiting from the South pole\footnote{This can
be realized by sending a flux of positive charges through the
North pole, and a flux of negative charges through the South
pole.}. This generates a certain stationary electromagnetic field.
The analysis in \cite{Damour:1978cg} of the regularity of the
field structure on the event horizon has shown that (even for a
non-rotating hole) the magnetic field generated by the current
{\it must} be accompanied by a correlated electric field. It was
then explicitly verified that the latter electric field is
responsible for generating an electric potential difference ( an
``EMF'') between the two poles such that Ohms' law $ V = R I$ is
satisfied, with a resistance $R \sim 30$ Ohms  computable ``as
if'' the black hole horizon were a conducting surface of
resistivity equal to 377 Ohms (i.e. $4 \pi$). This electric
potential difference is, in principle, observable, and might
actually play a significant role in magnetic-field based
mechanisms for extracting energy from black holes
\cite{Ruffini:1975ne,Blandford:1977ds}, which are believed to be
important in active galactic nuclei and other astrophysical
processes.

If we pass a current $I$ through a wormhole, we expect, at face
value, to generate only a magnetic field. More precisely, adopting
the geometry of current injection of  \cite{Damour:1978cg}, and
solving Maxwell equations for a general spherically symmetric
wormhole background, we find a solution involving a purely
magnetic field strength \be \lb{WH} F^{\rm wormhole} = dA = {2 I
\over \sin \theta} \sqrt{B \over A} d \theta \wedge d r ~~.\ee
This is consistent with \cite{Damour:1978cg}, but represents only
the {\it magnetic} part of the result

\be \lb{BH} F^{\rm blackhole} = dA = {2 I \over \sin \theta} d
\theta \wedge ( dt + { dr \over 1- 2GM/r})~~. \ee

If that were all, the difference between the last two results
would signal a clear physical distinction between a wormhole and a
black hole. However, (\ref{WH}) has been derived by looking for a
{\it stationary} solution of Maxwell equations in a wormhole
background. This would be reasonable for a usual object through
which passes a stationary current. But a wormhole is not a usual
object, and we must take into account the crucial physical fact
which played an important role above. When $\lambda$ is very
small, one must remember that far from leading to a stationary
state the charges (of opposite signs) continuously sent onto the
poles of a wormhole will appear (on usual external timescales) to
accumulate on the North and South poles of the $r=2GM$ throat,
though the effect of these localized accumulated charges will tend
to uniformly spread out into a low-multipole electric field. When
thinking more about this rather complicated problem, one then
realizes that we can use the same arguments that we have used
already above. It consists essentially in saying that the retarded
Green function (now considered for a Maxwell field) in a wormhole
spacetime will precisely mimic (if $\lambda$ is small enough)
 the black-hole retarded Green function
if one considers a source which started falling in a finite time
ago, and an observer having also a finite observing time window.

\section{ Quantum effects}
\setcounter{equation}0
 Let us finally consider  quantum
effects in a wormhole metric, and compare them to the black hole
case.

We have already seen that in a wormhole spacetime there are
classical geodesics, absent in the Schwarzschild metric, which
oscillate around the throat region. We therefore expect that there
will be corresponding quantum modes which are trapped within the
throat. Actually, we have already mentioned them above. Indeed,
when considering, for simplicity, a quantum scalar field $\Phi$
propagating in a wormhole metric, we can decompose it in modes of
frequency $\omega$ and angular momentum $(l,m)$. This leads to the
separated radial equation written in equation (\ref{8}) above.

The effective radial potential (\ref{9}) reaches a minimum
(positive) value  ${\rm min} \ U={\lambda^2\over 4G^2M^2}[{1\over
2}+l(l+1)]$ at the center of the throat $r=2GM$. Note that this
minimum value is positive, but tends to zero like the square of
$\lambda$. This minimum is surrounded on both sides by much higher
positive maxima located around (we take the limit $l\gg 1$ in which
these expressions simplify) $r={3GM\over 1+\lambda^2}$ and
of value
$${\rm max}\ U={1\over 27G^2M^2}(l(l+1)+{2\over 3})+{1\over
9G^2M^2}(l(l+1)+{5\over 9})\lambda^2~~.$$

 As we already mentioned above when discussing QNM modes, the
radial equation (\ref{8}) admits a discrete set of long-lived
resonances within this potential well. The lowest (nearly real)
energy levels can be obtained by looking in the throat region $ r
- 2GM \ll 2GM$. There the relation between the coordinates $r$ and
$z$ is given by $r(z)=2GM(1+\lambda^2\sinh^2{z\over 4GM})$. The
effective radial potential takes the form (for arbitrary $l$)
$$
U(z)={\lambda^2\over 8G^2M^2}(1+2l(l+1))+{\lambda^2\over
4G^2M^2}(1+l(l+1))\sinh^2({z\over 4GM})~~.
$$
The discrete spectrum for this potential can be obtained in the
WKB approximation  \cite{Solodukhin:2005qy},
\be
\omega_n={\pi\over 8GM \ln({1/\tilde{\lambda}})}(n+{1\over 2})\, ,
\, n\in {Z} \lb{10} \ee
where we defined
$\tilde{\lambda}^2=\lambda^2 (1+l(l+1))$ and  neglected terms of
order $\lambda^2$ (and $\tilde{\lambda}^2$).

For $l=0$ the first discrete  level appears far above the bottom
of the potential well ${\lambda^2\over 8G^2M^2}$ but far below the
top of the potential. Thus, there is a large number ${\cal N}_0$
of discrete levels inside this $l=0$ well. This number can be
estimated as
\be
 {\cal N}_0\simeq {\sqrt{27}\over 4\pi} \ln{1\over \lambda}~~.
\lb{11}\ee
If we consider the case where $\lambda$ is given by
(\ref{2}), we note that the number of bound states is of the order
of
 ${\cal N}_0\sim GM^2$, which is of the same order as the entropy of the
Schwarzschild black hole. Note, however, that this was only the
$l=0$ modes. If we consider $l \neq 0$ modes we will have a
similar spectrum of quasi-bound states, and the number of bound
states for given values of $l$ and $m$ will be of order (for large
$l$ and neglecting $\lambda^2$ terms) \be{\cal N}_{lm}\simeq
{8\over \pi \sqrt{27}}\sqrt{l(l+1)+{2\over 3}}~ \ln(1/
\tilde{\lambda})\sim \sqrt{l(l+1)+{2\over 3}}~GM^2~~. \lb{12}\ee

 It would
be interesting to study more carefully whether these
quasi-bound states could be considered as analogs of the somewhat
mysterious ``black hole microstates'' which are supposed to be
counted by the Bekenstein-Hawking entropy (see e.g.
\cite{Damour:2004kw} for a review). Even more interestingly, as
the wormhole resonances discussed here are all unstable, it is
tempting to conjecture that they might somehow mimic the Hawking
radiation. We have in mind here a mechanism of the following sort.
During the collapse leading to the (assumed) formation of a
wormhole, the quantum field $\Phi$ will be left in a state where
many of the wormhole resonance modes will be excited. The modes
which have a large decay width $\Gamma = - \Im \omega$ (like the
ordinary QNM modes) will be radiated quite fast. But the modes
which are deep down within the double-humped potential well will
have a
 very small decay width $\Gamma $ and will slowly trickle out
 of the potential well, thereby generating a nearly continuous
 radiation emitted by the wormhole.

 However, though we anticipate that this ``wormhole radiation''
 might (especially for the choice (\ref{2}) of $\lambda$)
 coarsely model Hawking's radiation, we do not think that it
 will be possible to reproduce with  precision the
 specific thermal-like grey-body spectrum predicted in \cite{Hawking:1974sw}.
 Indeed, this spectrum is a delicate consequence of the fact that
 the modes of a quantum field which ``straddle'' the event horizon
 get torn into two outgoing modes, one of which exits at radial
 infinity in the form of a quasi-thermal spectrum.

\section{ Conclusions}
\setcounter{equation}0
 In conclusion we considered a wormhole
spacetime as a ``foil'' to a Schwarzschild black hole, to learn to
what extent the  observational features of a black hole do really
depend on the presence of an event horizon. Indeed,   unlike a
black hole, a wormhole geometry is globally static and does not
have an event horizon. It was clear from the start that, as the two
spacetimes have a nearly identical geometry for $r > 2GM$, they
would have very similar closed geodesics, and  would therefore be
practically indistinguishable in astronomical observations that
depend only on the external gravitational field. However, and more
surprisingly, we found that many observational features that were
thought to crucially depend on the presence of an event horizon
were well mimicked by a wormhole, if the parameter $\lambda$ is
sufficiently (exponentially) small. This includes, the apparently
irreversible accretion of matter down a hole, no-hair properties,
quasi-normal-mode ringing, and even the dissipative properties of
black hole horizons, such as a finite surface resistivity equal to
377 Ohms  \cite{Damour:1978cg}.

Finally, we conclude that the only ways to observationally
distinguish a wormhole from a black hole are: (1) either to
observe classical phenomena (such as matter accretion) over the
long ``wormhole bounce'' time scale  $\Delta t= 2GM \ln{1\over
\lambda^2}$ ,  (2) or to observe the Hawking radiation presently
coming out of a hole. Interestingly, there is a link between these
two methods: when $\lambda$ takes the value (\ref{2}) suggested in
\cite{Solodukhin:2005qy} the classical wormhole bounce time scale
becomes comparable to the quantum evaporation time of a
Schwarzschild black hole $\Delta t=16\pi G^2M^3 $. However, in the
case where $\lambda$ takes the value (\ref{2}) both methods are
unpractical because the time scale $\Delta t=16\pi G^2M^3 $ is
much too large for usual astrophysical masses, and/or the Hawking
temperature is much too low (being much smaller than the 3 $K$
cosmological background).

It remains interesting to keep in mind that most of the
phenomenology of black holes does not really depend on the
presence of an horizon, and also (though this deserves more study)
that a wormhole could somehow mimic the Hawking radiation, as well
as may provide a simple way of visualizing the microstates storing
the information apparently ``lost'' during a gravitational
collapse. One would, however, need a more detailed model of the
{\it formation} of a wormhole to address this issue.

\bigskip

\noindent{\bf Acknowledgments:} \noindent SS would like to thank
V. Frolov, T. Jacobson and D. Lowe for useful conversations at
earlier stages of this work. SS is grateful to the IH\'ES
 for the hospitality extended to him while this work was in progress.



\end{document}